\documentclass[conference]{IEEEtran}
\usepackage[latin9,utf8]{inputenc}
\usepackage{amsmath}
\usepackage{graphicx}
\usepackage{multirow}
\usepackage{algorithm}
\usepackage{algpseudocode}
\usepackage{longtable}
\newcommand\scalemath[2]{\scalebox{#1}{\mbox{\ensuremath{\displaystyle #2}}}}
\makeatletter

\def\ps@IEEEtitlepagestyle{
  \def\@oddfoot{\mycopyrightnotice}
  \def\@evenfoot{}
}
\def\mycopyrightnotice{
  {\footnotesize 979-8-3503-4602-2/22/\$31.00~\copyright~2022 IEEE\hfill} 
  \gdef\mycopyrightnotice{}
}

\@ifundefined{showcaptionsetup}{}{
 \PassOptionsToPackage{caption=false}{subfig}}
\usepackage{subfig}
\makeatother

\usepackage{eso-pic}
\newcommand\AtPageUpperMyright[1]{\AtPageUpperLeft{
 \put(\LenToUnit{0.5\paperwidth},\LenToUnit{-1cm}){
     \parbox{0.5\textwidth}{\raggedleft\fontsize{9}{11}\selectfont #1}}
 }}
\newcommand{\conf}[1]{
\AddToShipoutPictureBG*{
\AtPageUpperMyright{#1}
}
}

\begin{document}

\title{Brain Tumor Segmentation using Enhanced U-Net Model with Empirical Analysis}
\conf{2022 25th International Conference on Computer and Information Technology (ICCIT)} 

\author{\IEEEauthorblockN{MD Abdullah Al Nasim\IEEEauthorrefmark{1}, Abdullah Al Munem\IEEEauthorrefmark{1}, Maksuda Islam\IEEEauthorrefmark{1}, \\ Md Aminul Haque Palash\IEEEauthorrefmark{1}, MD. Mahim Anjum Haque\IEEEauthorrefmark{1}, and Faisal Muhammad Shah\IEEEauthorrefmark{2}}
\IEEEauthorblockA{\IEEEauthorrefmark{1}Dept of Research and Development, Pioneer Alpha}
\IEEEauthorblockA{\IEEEauthorrefmark{2}Dept of Computer Science and Engineering, AUST}
\IEEEauthorblockA{Email:  nasim.abdullah@ieee.org,  abdullahalmunem@gmail.com,  maksudalima321@gmail.com, \\ aminulpalash506@gmail.com, mahim@vt.edu, faisal.cse@aust.edu}
}

\maketitle

\begin{abstract}
Cancer of the brain is deadly and requires careful surgical segmentation. The brain tumors were segmented using U-Net using a Convolutional Neural Network (CNN). When looking for overlaps of necrotic, edematous, growing, and healthy tissue, it might be hard to get relevant information from the images. The 2D U-Net network was improved and trained with the BraTS datasets to find these four areas. U-Net can set up many encoder and decoder routes that can be used to get information from images that can be used in different ways. To reduce computational time, we use image segmentation to exclude insignificant background details. Experiments on the BraTS datasets show that our proposed model for segmenting brain tumors from MRI (MRI) works well. In this study, we demonstrate that the BraTS datasets for 2017, 2018, 2019, and 2020 do not significantly differ from the BraTS 2019 dataset's attained dice scores of 0.8717 (necrotic), 0.9506 (edema), and 0.9427 (enhancing). 
\end{abstract}


\section{Introduction}
The area of medical imaging has undergone a revolution in recent decades with the application of machine learning and deep learning algorithms for a tumor or other disease segmentation, identification, and survival prediction. In addition, it assists doctors in making an early diagnosis of brain malignancies to improve foretelling.
Adults most frequently develop gliomas, which are thought to have glial cell origins and invade adjacent tissues. Primary brain tumors are gliomas. High-grade glioblastoma (HGG) and low-grade glioblastoma (LGG) are two more subtypes of gliomas. While magnetic resonance imaging (MRI) modalities have been manually examined by radiologists to generate quantitative information, segmenting 3D modalities is cumbersome, with variations and errors. This challenge is further increased if tumors vary in size, shape, and placement\cite{du2020medical}.

Enhancing disease diagnosis, cure planning, monitoring, and clinical ordeals depends heavily on segmenting brain tumors from neuroimaging modalities. To determine the spot and size of the tumor, accurate brain tumor segmentation is necessary. These tumors can develop in practically any area and come in various sizes and shapes. Furthermore, the rigorousness of a tumor may coincide with the vitality of vital brain tissue, and they are typically poorly contrasted. As a result, it is challenging to differentiate fit tissue from a tumor. For example, while T1c has a bright tumor border, T2w has a bright tumor area. On the other hand, the FLAIR scan aids in isolating the edema from the cerebrospinal fluid (CSF)\cite{bauer2013survey}.
Integrating data from various MRI modalities, such as T1-weighted MRI (T1), T1-weighted MRI with contrast (T1c), T2-weighted MRI (T2), and Fluid-Attenuated Inversion Recovery (FLAIR), is a fundamental approach to address this problem. 

Automated and semi-automatic segmentation techniques have allowed experts to work more efficiently. However, since tumorous cells can arise anywhere inside the brain tissues and can vary in size, appearance, and shape, automatically segmenting a brain tumor and its sub-regions is difficult\cite{al2020deep}. Nevertheless, these imaging technologies' ability to accurately and quickly segment brain tumors can help doctors treat tumors safely, especially during surgery, without endangering the brain's healthy areas\cite{raza2022dresu}.

Different architectures of CNNs have been used and are being used to do automatic segmentation of 3D MRI Images \cite{pereira2016brain}\cite{bauer2013survey}. Earlier, it was a difficult task due to the absence of medical Image data and the huge amount of processing power. However, challenges like Brain Tumor Segmentation(BraTS) made it easier by publishing annotated 3D MRI images. 



In this paper, we aim to make an extensive comparison of these different kinds of CNN models, along with proposing a 2D UNET model of our own to help doctors to improve their performance with the initial scan provided by an automated and intelligent system. We also compared the gained metrics' results on the BraTS 2017-2020 dataset using traditional ML models and other CNN models with our proposed one. Artificial Intelligence (AI) has been used in various domain such as COVID-19 \cite{nasim2021prominence} \cite{abdullah2021prominence}, Fine-Grained Image Generation \cite{palash2021fine}, nonuniform compressed sensing \cite{karim2019rl}, SPI-GAN \cite{karim2021spi} and many more. The humongous application of AI shows that it can be used to solve different problem in various domain.

\section{Related Work}

Classical machine learning techniques were used for segmentation earlier until deep learning techniques like convolutional neural network\cite{cciccek20163d} brought significantly advanced performance for automatic segmentation. To visualize the statistic, if the BraTS 2017 dataset is only considered, more than 55 papers were introduced based on the BraTS 2017 dataset. Based on those papers, approximately 42 models are based on CNN; among those CNN models, 16 models were inspired by U-Net or V-Net, 6 models were founded on DeepMedic, and others are founded on traditional machine learning models, for example, Random Forest (5 Models) and SVM (2 Models). Since this study focuses on deep learning techniques, this section highlighted some prior work on brain tumor segmentation. To predict the label of each pixel, Havaei et al. \cite{havaei2017brain} proposed a CNN model with two-pathway and, as input, used a regional picture block. ResNet used the effective bottleneck structure to obtain outstanding performance. The oldest and most well-known image semantic segmentation technique, "U-Net," was suggested by Ronneberger et al. \cite{cciccek20163d}. Since then, U-Net has been more well-liked and successfully used in various medical imaging and computational pathology applications. 

A LSTM Multi-modal UNet \cite{menze2014multimodal}, combined two components, multi-modal UNet, \& convolutional LSTM. They implemented both 3D \& 2D multi-modal UNet. 2D multi-modal UNet was implemented as a contrast. There are several UNet encoding paths and only one decoding path in 2D multi-modal UNet. On the BraTS 2015 dataset, they performed IOU on each class and compared UNET, and Multi-modal UNET, 3 classes out of 4 gave better results for multi-modal UNET. Even though their model yielded a better result, they mentioned that their complex model has a slow segmentation speed.

Agravat et al. \cite{agravat20203d} suggested a 3D encoder-decoder FCNN model. To portray the performance of their proposed model on segmentation, they used 4 different evaluation metrics on the 2020 BraTS dataset. They achieved 88.1\% dice similarity coefficient (DSC), 84.6\% Sensitivity,  06.508\% Hausdorff95, and 99.9\% Specificity for whole tumor (WT), 83.1\% DSC, 80.2\% Sensitivity,  07.275\% Hausdorff95, and 99.9\% Specificity for tumor core (TC), and 75.7\% DSC, 75.9\% Sensitivity,  31.531\% Hausdorff95, and 99.9\% Specificity for enhancing tumor (ET) on BraTS 2020 Validation dataset using mean statistic. Kumar Anand et al. \cite{anand2020brain} also proposed a 3D FCNN model, which gave a poor result for all 4 evaluation metrics compared to the previously mentioned Agravat et al. model. The authors of `Glioma Detection and Segmentation Using Deep Learning Architectures'\cite{gomathi2022glioma} literature LeNET and AlexNET methods. Using ground truth samples, they applied their methods to BraTS 2019 and 2020 datasets for evaluating metrics (Sensitivity, Specificity, Accuracy). Obtained results for LeNET in this paper were 94.56\% of Sensitivity, 95.32\% of Specificity, and 96.53\% Accuracy. In this paper, the Alex NET results were 94.53\% of Sensitivity, 95.32\% of Specificity, and 97.88\% of Accuracy. A few authors also proposed RCNN to solve the brain tumor segmentation problem \cite{singh2022novel} \cite{zhuge2020automated} \cite{khairandish2020hybrid}. Among them, Singh et al. \cite{singh2022novel} trained their model using the BraTS 2020 dataset and got a precision of 0.79, recall of 0.72, and dice coefficient of 0.75. 
 
Xu et al. \cite{xu2019lstm} considered the correlations between the modalities of the BraTS dataset and exploit the in-depth information. They used LSTM multi-modal 2D U-Net on BraTS 2015 dataset to experiment. They used LSTM to exploit sequential information from the images and Dense connections to obtain the correlations between the modalities. They used a feature size of 64 in the first convolution layer, and the maximum feature size of the encoder path is 256. They achieved dice scores of 0.7309, 0.6235, and 0.4254, Sensitivity of 0.6376,0.5975, and 0.7163, PPV of 0.8979,0.7264, and 0.3860 for whole, core and enhancing tumor respectively.
 
There are some other related studies where the authors used simple U-Net architecture with different numbers of encoder-decoder blocks and feature sizes. There are some other significant work done by the authors of \cite{shah2019brain} \cite{hossain2019brain}, where they used CNN to segment brain tumor on medical images. Dong et al. \cite{dong2017automatic} utilized 2D U-Net architecture on the BraTS 2015 dataset and gained dice scores of 0.86, 0.86, and 0.65 for whole, core, and enhancing tumors, respectively. In paper \cite{islam2019brain}, the authors integrated a 3D attention module of the decoder paths of simple U-Net architecture and achieved mean dice scores of 0.704, 0.898, and 0.792, mean sensitivity of 00.751, 0.900, and 0.816, mean specificity of 0.998, 0.994, and 0.996, mean hausdorff95 of 7.05, 6.29, and 8.76 for enhancing tumor, whole, and core respectively on BraTS 2019 Dataset.

\section{Dataset Description}
The brain tumor segmentation(BraTS) challenge was first held in 2012 with the aim of uniting the research communities worldwide. Since then, this competition has taken place yearly as a part of the MICCAI conference till 2020. The data were accumulated from numerous institutions employing different MRI scanners. The organizer delivered these data after basic pre-processing, co-registered to the exact anatomical template, incorporated to the same resolution, $1 mm^3$, and skull stripped\cite{bakas2018identifying}. The provided datasets for 2017-20 were slightly different compared to 2012-16's. However, from BraTS 2017-2020, they publish similar datasets with the same modalities; all are pre-operative data. 

This study uses the BraTS 2017 - 2020 dataset to train, evaluate and compare the models. These datasets contain 3D MRI brain scans for a precise type of brain tumor, Glioma. Table \ref{t1} holds the detailed information for the BraTS 2017, 2018, 2019 and 2020 datasets. All of them have separate training, validation, and testing dataset. Expert physicians from different institutions did the annotation. All of these were provided to make a similar task, Segmentation \& Survival prediction, and all of the datasets are pre-operative MRIs.

These datasets consist of 4 distinct MRI modalities, namely, native (T1), T2-weighted (T2), post-contrast T1-weighted (T1ce), and fluid-attenuated inversion recovery (FLAIR). Each MRI modality has 155 slices per volume. The enhancing tumor, the necrotic, the peritumoral edema, and the non-enhancing tumor core are the three tumor sub-regions that are annotated. 1 for necrosis and a non-enhancing tumor (NCR/NET), 2 for peritumoral edema (ED), 4 for an enhancing tumor, and 0 for background in each training patient's annotated labels. Integrating three regions—the whole tumor (also known as Whole Tumor or Whole: labels 1, 2, and 4), the tumor core (also known as Core: labels 1 and 4), and the enhanced tumor ( known as Enhancing Tumor or Enhancing: label 4). 

\begin{table}[htbp]
\caption{Description of the BraTS dataset\cite{bakas2018identifying}\cite{ahmad2020context}}
\label{t1}
\begin{tabular}{|c|c|c|c|c|}
\hline
Dataset                     & Type       & Patients(No of Cases) & HGG                  & LGG                 \\ \hline
\multirow{2}{*}{BraTS 2017} & Training   & 285                   & \multirow{2}{*}{210} & \multirow{2}{*}{75} \\ \cline{2-3}
                            & Validation & 46                    &                      &                     \\ \hline
\multirow{2}{*}{BraTS 2018} & Training   & 285                   & \multirow{2}{*}{210} & \multirow{2}{*}{75} \\ \cline{2-3}
                            & Validation & 66                    &                      &                     \\ \hline
\multirow{2}{*}{BraTS 2019} & Training   & 335                   & \multirow{2}{*}{259} & \multirow{2}{*}{76} \\ \cline{2-3}
                            & Validation & 125                   &                      &                     \\ \hline
\multirow{2}{*}{BraTS 2020} & Training   & 369                   & \multirow{2}{*}{293} & \multirow{2}{*}{76} \\ \cline{2-3}
                            & Validation & 125                   &                      &                     \\ \hline
\end{tabular}
\end{table}

\section{Methodology}
\begin{figure}[ht]
\centering
\includegraphics[width=0.3\textwidth]{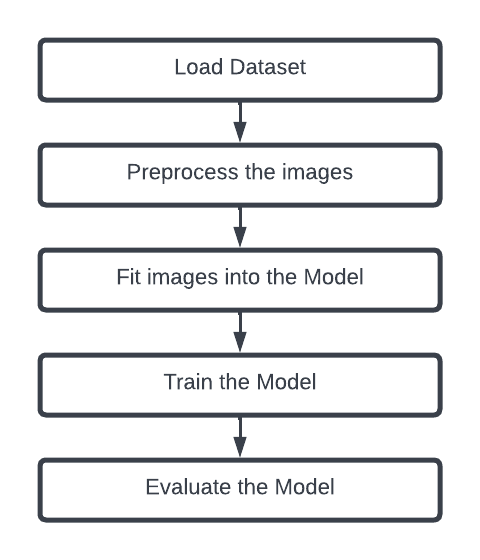}
\caption{Workflow of proposed model}
\label{fig:1}
\end{figure}

The 2D U-Net architecture of CNN has been employed for brain tumor segmentation. Each dataset (BraTS 2017, 2018, 2019, and 2020) has used some preprocess technique. A data generator class has been used in this experiment to fit the dataset into the main memory. After that, the dataset was divided into train, test, and validation datasets and fed into the U-Net model. Then, we evaluate the trained model and measure the performance based on different metrics. Figure \ref{fig:1} shows the workflow of the model.

\begin{algorithm}
\caption{Algorithm of U-Net Approach with pre-processing}
\begin{algorithmic}
\For{each image in the dataset}
\State  Resize the image into dimensions of 128 * 128 * 3
\State  Slice the Image to remove blank portions
\State  One Hot encode the SEGMENT CLASSES
\State	Normalize the input image
\EndFor
\State Train U-Net with 235 iterations
\State Evaluate the Model
\end{algorithmic}
\end{algorithm}

\begin{algorithm}
\caption{Algorithm for the Evaluation}
\begin{algorithmic}
\State loadDataset()
\State DataGenerator()
\State train\textunderscore test\textunderscore split()
\State loadModel()
\For{each epoch in epochNumber}
\For  {each batch in batchSize}
\State		y\textunderscore predict = model(train\textunderscore feature\textunderscore generator)
\State		loss = categorical\textunderscore crossentropy(y\textunderscore actual, y\textunderscore predict)
\State		adamOptimizer(loss, learning\textunderscore rate)
\State		evaluation()
\EndFor
\EndFor
\State return
\end{algorithmic}
\end{algorithm}

\begin{figure*}[ht]
\centering
\includegraphics[width=0.8\textwidth]{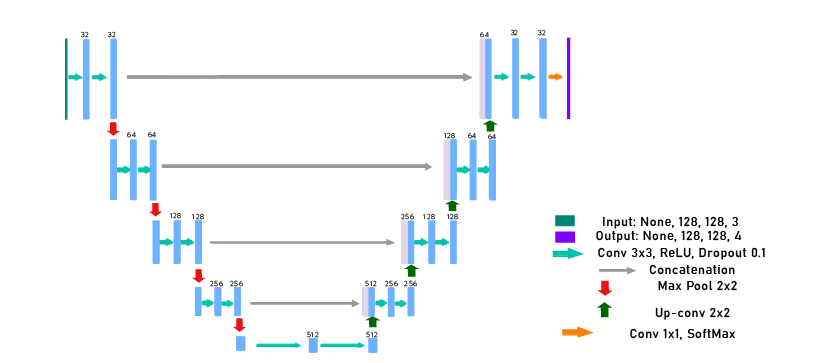}
\caption{U-net architecture for our working methodology}
\label{fig:2}
\end{figure*}

\subsection{Network Architecture}\label{na}
Figure \ref{fig:2} depicts the 2D U-Net network architecture. The expanding path on the right side of figure \ref{fig:2} and the contracting path on the left are the two main paths of it.
The typical convolutional network architecture constitutes the contracting path. It consists of two 3x3 convolutions and an activation function, namely, rectified linear unit (ReLU). To downsample, it uses a 2x2 max pooling operation. The number of features doubles with each down-sampling step. 
2x2 transposed convolutional layers constituted the expansive path, which is used for upsampling the feature map and concatenation with the affiliated feature map of the contracting path, and two 3x3 convolutions with ReLU as an activation function. At each step of upsampling, the number of features is divided by 2. At the last layer, a 1x1 convolution has been used for mapping each feature vector to the wanted number of classes.

\subsection{Data Pre-processing}\label{preprocess}
The datasets contain 4 distinct MRI modalities. Training the CNN model with all those modalities is computationally expensive. Thus, FLAIR and T1ce are considered as input for the 2D U-net. FLAIR and T1ce contain most of the valuable information in this dataset. But each MRI modality contains unnecessary black background, which is not required for the training phase; rather, it increases the computational time. For this reason, the images are sliced into several regions (100 slices for 2D U-net) to minimize the background and specify the informative portion of the images. 
Both input and output images are resized to the required shape of the U-net model. And lastly, the input and output images are normalized using MinMaxScaler because the intensity of each channel is $2^{16}$ bits. For the Segmented images, one hot encoding has been used. 
After that, we split the dataset into train, validation, and test datasets with a ratio of 68\%, 20\%, and 12\% of the whole dataset, respectively. 

\begin{figure}[ht]
\centering
\includegraphics[width=0.5\textwidth, height=8cm,keepaspectratio]{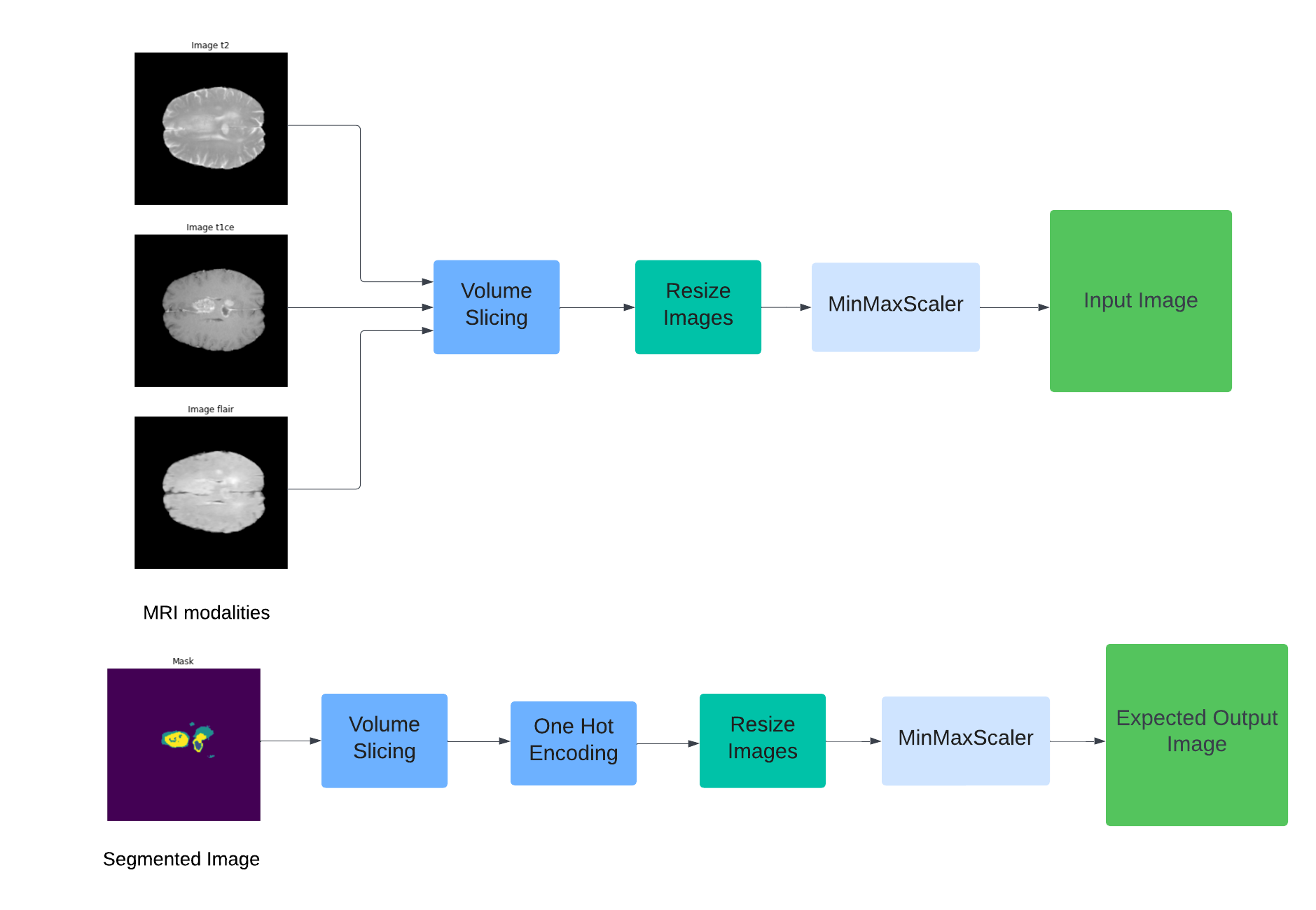}
\caption{Pre-processing steps }
\label{fig:3}
\end{figure}

\subsection{Experimental Setup}
The model was created using Keras and Tensorflow. The experimentations performed in Kaggle and Google Colaboratory contain Python version 3.7.12 and Tensorflow version 2.6.4. The chosen hyperparameters shown in table \ref{t2}.

\begin{table}[]
\caption{Hyperparameter Values of the proposed model.}
\label{t2}
\begin{tabular}{|l|l|}
\hline
Number of epochs   & 235                                                                                                                                                                                                                  \\ \hline
Batch size         & 1                                                                                                                                                                                                                    \\ \hline
Optimizer          & Adam                                                                                                                                                                                                                 \\ \hline
Loss               & Categorical cross-entropy                                                                                                                                                                                            \\ \hline
Metrics            & \begin{tabular}[c]{@{}l@{}}Accuracy, MeanIoU, Dice coefficient, \\ Precision, Sensitivity, Specificity, \\ Dice coefficient for necrotic, \\ Dice coefficient for edema, \\ Dice coefficient for enhancing\end{tabular} \\ \hline
Callback functions & Early stopping, Csv logger, Model Checkpoint                                                                                                                                                                         \\ \hline
\end{tabular}
\end{table}

\subsection{Evaluation Metrics}
Numerous evaluation criteria are calculated to check the efficiency of the implemented models. The dice coefficient is mostly used for evaluating medical image segmentation. In this experiment, dice coefficient has been computed for each class (necrotic Eq. \ref{dcn}, edema Eq. \ref{dce}, enhancing Eq. \ref{dcen}) and the predicted segmentation. As well as accuracy, MeanIoU, Precision, Sensitivity, and Specificity are computed to evaluate overall model performance and genericity. For each evaluation criterion, a higher score means better segmentation performance.

True Positive (TP) in the equations denotes a tumor match between the anticipated and actual tumor. True Negative (TN) indicates a match between the expected and actual non-tumor. False Positive (FP) denotes that the predicted tumor area is not the actual tumor, and False Negative (FN) means that the expected non-tumor is the actual tumor area.

\begin{align} \label{iou}
\scalemath{0.7}{
     IoU = \frac{Area of Intersection (TP)}{Area of Union (TP + FP + FN)}} 	
\end{align}

\begin{equation} \label{dc}
  \scalemath{0.7}{
    Dice Coefficient = \\
    \frac{2 Intersection (TP)}{Intersection + Union (TP + FP + FN)}}		
\end{equation}

\begin{equation} \label{dcn}
  \scalemath{0.7}{
    Dice Coefficient Necrotic= \\
    \frac{2 Intersection of Necrotic (TP)}{Union of Necrotic (TP + FP + FN)}}	
\end{equation}

\begin{equation} \label{dce}
  \scalemath{0.7}{
    Dice Coefficient Edema=\\
    \frac{2 Intersection of Edema (TP)}{Union of Edema (TP + FP + FN)}}	
\end{equation}

\begin{equation} \label{dcen}
  \scalemath{0.7}{
    Dice Coefficient Enhancing=\\
    \frac{2 Intersection of Enhancing (TP)}{Union of Enhancing (TP + FP + FN)}}
\end{equation}
\section{Experimental Result And Comparison}
The performance of the 2D U-Net model on 4 different datasets (BRAST 2017, 2018, 2019, 2020) has been compared in this experiment. The proposed 2D U-Net outperformed on BRAST dataset compared to other CNN-based models such as FCNN and RCNN. This model gives the most robust result on BRAST 2019 dataset compared to the other BRAST datasets. Since the difference between the BRAST datasets is almost negligible, thus this model can be used on any BRAST dataset. The result of this experiment is shown in Table \ref{t3}.

\begin{table}[]
\caption{Comparison With Previous Works}
\label{t4}
\begin{tabular}{|c|c|ccc|}
\hline
\multirow{2}{*}{Model}                                                     & \multirow{2}{*}{\begin{tabular}[c]{@{}l@{}}Dataset \\ (BraTS)\end{tabular}} & \multicolumn{3}{c|}{Dice Score}                                                                                                                                             \\ \cline{3-5} 
                                                                           &                                 & \multicolumn{1}{c|}{\begin{tabular}[c]{@{}c@{}}Necrotic/\\ Core\end{tabular}} & \multicolumn{1}{c|}{\begin{tabular}[c]{@{}c@{}}Edema/\\ Whole\end{tabular}} & Enhancing     \\ \hline
SVM \cite{bauer2011fully}                                                               & 2012                            & \multicolumn{1}{c|}{0.61}                                                     & \multicolumn{1}{c|}{0.71}                                                   & 0.73          \\ \hline
Random Forest \cite{bauer2012segmentation}                                                     & 2012                            & \multicolumn{1}{c|}{0.73}                                                     & \multicolumn{1}{c|}{0.59}                                                   & -             \\ \hline
Random Forest \cite{zikic2014segmentation}                                                     & 2013                            & \multicolumn{1}{c|}{0.70}                                                     & \multicolumn{1}{c|}{0.76}                                                   & 0.67          \\ \hline
2D CNN \cite{zikic2014segmentation}                                                            & 2013                            & \multicolumn{1}{c|}{0.73}                                                     & \multicolumn{1}{c|}{0.83}                                                   & 0.69          \\ \hline
3D CNN \cite{kamnitsas2016deepmedic}                                                            & 2015                            & \multicolumn{1}{c|}{0.73}                                                     & \multicolumn{1}{c|}{0.87}                                                   & 0.77          \\ \hline
\begin{tabular}[c]{@{}c@{}}3D encoder-decoder\\ FCNN \cite{agravat20203d}\end{tabular} & 2020                            & \multicolumn{1}{c|}{0.83}                                            & \multicolumn{1}{c|}{0.88}                                          & 0.78 \\ \hline
3D FCNN \cite{anand2020brain}                                                           & 2020                            & \multicolumn{1}{c|}{0.74}                                                     & \multicolumn{1}{c|}{0.88}                                          & 0.71          \\ \hline

The Proposed Model                                                           & 2019                            & \multicolumn{1}{c|}{\textbf{0.871}}                                                     & \multicolumn{1}{c|}{\textbf{0.95}}                                          & \textbf{0.94}          \\ \hline

\end{tabular}
\end{table}

\begin{figure}[ht]
\centering
\includegraphics[width=0.6\textwidth,width=10cm,height=6cm,keepaspectratio]{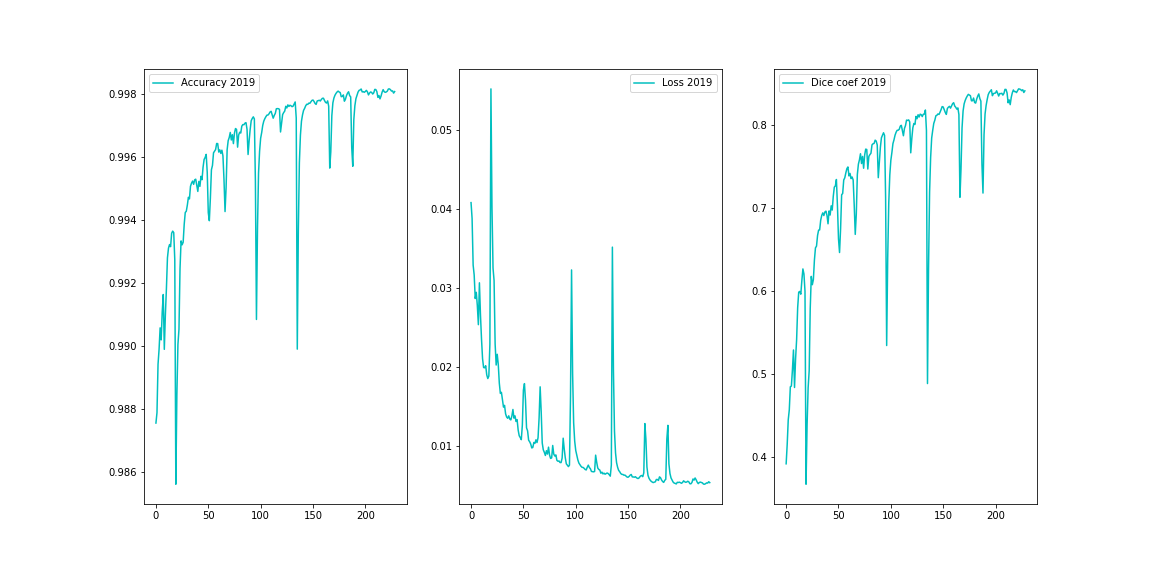}
\caption{The loss, accuracy, and dice score of BRAST 2019 dataset}
\label{fig:4}
\end{figure}

\begin{figure}[ht]
\centering
\includegraphics[width=0.6\textwidth,width=10cm,height=6cm,keepaspectratio]{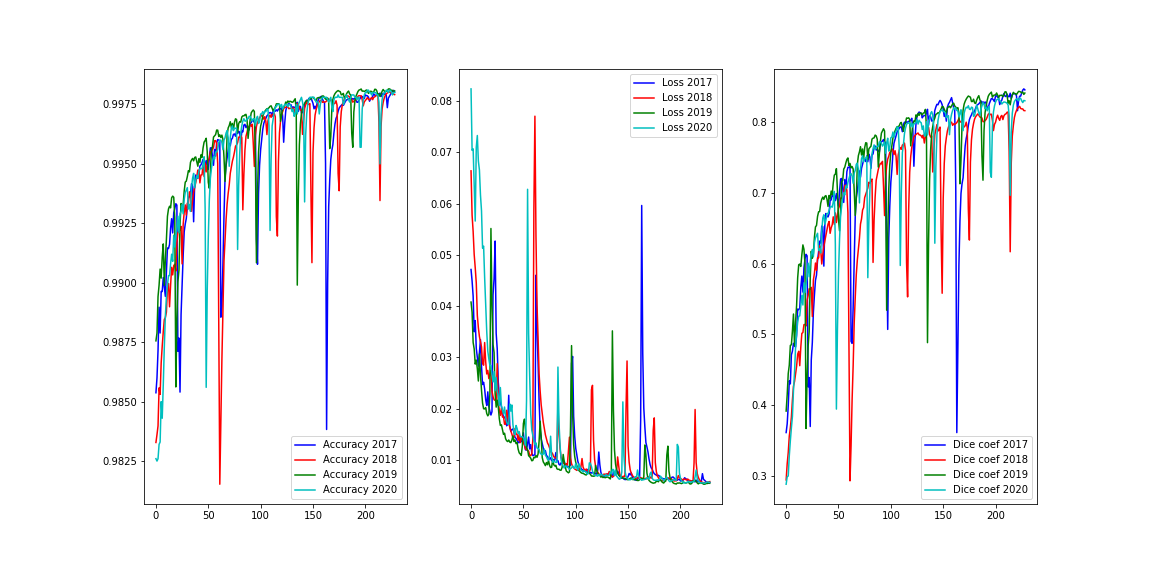}
\caption{The loss, accuracy, and dice score comparison of BRAST 2017, 2018, 2019, 2020 dataset}
\label{fig:5}
\end{figure}

Table \ref{t3} shows that the proposed model achieved highest Accuracy (0.9981), Mean IoU (0.9130), Precision (0.9974), Sensitivity (0.9971), Specificity (0.9991), Dice Score (0.8409) and minimal Loss (0.0054) on BRAST 2019. The highest Dice Scores of Edema, and Enhancing 0.9545, and 0.9490, respectively, achieved on the BraTS 2017 dataset. The highest Dice Scores of 0.8846 for Necrotic achieved on the BraTS 2020 dataset. The difference between the obtained result on these datasets are very less. It shows that the difference between those BRAST datasets is insignificant.  The loss, accuracy, and dice score during the training on BraTS datasets are shown in figure \ref{fig:4} and figure \ref{fig:5}. 

There are some other CNN-based models and traditional machine learning (ML) models have been used for segmenting Brain Tumors and medical images. The results of those models from the previous works are shown in table \ref{t4}, \ref{t5}.

\begin{table}[]
\caption{Comparison With CNN-based Model}
\label{t5}
\begin{tabular}{|c|c|c|c|c|}
\hline
Model                                                                          & \begin{tabular}[c]{@{}l@{}}Dataset\\ (BraTS)\end{tabular} & Accuracy       & \begin{tabular}[c]{@{}c@{}}Sensitivity/\\ Recall\end{tabular} & Specificity \\ \hline
R-CNN \cite{singh2022novel}                                                                  & 2020           & -              & .72                                                           & -           \\ \hline
LeNET \cite{gomathi2022glioma}                                                                 & 2019           & 0.944          & 0.956                                              & 0.945       \\ \hline
AlexNET \cite{gomathi2022glioma}                                                               & 2019           & 0.961          & 0.952                                                         & .951        \\ \hline
R-CNN \cite{zhuge2020automated}                                                                 & 2018           & 0.963          & 0.935                                                         & 0.972       \\ \hline
\begin{tabular}[c]{@{}c@{}}Hybrid Fast \\ R-CNN\\ \& SVM \cite{khairandish2020hybrid}\end{tabular} & 2015           & 0.987 & -                                                             & -           \\ \hline
The proposed model                                                                 & 2019           & \textbf{0.9981}          & \textbf{0.9971}                                                         & \textbf{0.9991}       \\ \hline
\end{tabular}
\end{table}

\begin{table*}[h]
\caption{Experimental Result on BraTS 2017- 2020 dataset with our approach}
\label{t3}
\begin{tabular}{|l|l|l|l|l|l|l|l|l|l|l|}
\hline
\textbf{Dataset}    & \textbf{Loss}   & \textbf{Accuracy} & \textbf{Mean IoU} & \textbf{Precision} & \textbf{Sensitivity} & \textbf{Specificity} & \textbf{Dice Score} & \textbf{\begin{tabular}[c]{@{}l@{}}Dice Score\\ (necrotic/\\ core)\end{tabular}} & \textbf{\begin{tabular}[c]{@{}l@{}}Dice Score \\ (edema)\end{tabular}} & \textbf{\begin{tabular}[c]{@{}l@{}}Dice Score \\ (enhancing)\end{tabular}} \\ \hline
\textbf{BraTS 2017} & 0.0056          & 0.9980            & 0.9637            & 0.9973             & 0.9970               & 0.9972     & 0.8453              & 0.8782                                                               & \textbf{0.9545}                                                        & \textbf{0.9490}                                                            \\ \hline
\textbf{BraTS 2018} & 0.0057          & 0.9979            & 0.8927            & 0.9972             & 0.9970               & 0.9940      & 0.8160              & 0.8843                                                                         & 0.9368                                                                 & 0.8348                                                                     \\ \hline
\textbf{BraTS 2019} & \textbf{0.0054} & \textbf{0.9981}   & \textbf{0.9130}   & \textbf{0.9974}    & \textbf{0.9971}      & \textbf{0.9991}      & \textbf{0.8409}     & 0.8717                                                                         & 0.9506                                                                 & 0.9427                                                                     \\ \hline
\textbf{BraTS 2020} & 0.0056          & 0.9980            & 0.8935            & 0.9973             & 0.9970               & 0.9983      & 0.8300              & \textbf{0.8846}                                                                         & 0.9478                                                                 & 0.8544                                                                     \\ \hline
\end{tabular}
\end{table*}

Table \ref{t4} shows that the 3D encoder-decoder F-CNN model of paper \cite{agravat20203d} achieved the highest dice score (0.83 for core, 0.88 for whole and 0.78 for enhancing) among other models, including traditional ML models. Table \ref{t5} shows that The Hybrid Fast R-CNN and SVM Model of paper \cite{khairandish2020hybrid} achieved the highest accuracy of 0.987. The R-CNN model of paper \cite{singh2022novel} achieved the highest precision of 0.79, although the authors of the other papers didn't mention the precision in their papers. The R-CNN model of paper \cite{zhuge2020automated} achieved the highest specificity of 0.972, and the LeNET model of paper \cite{gomathi2022glioma} achieved the highest Sensitivity/ Recall of 0.956. From the table \ref{t4}, \ref{t5}, it can be concluded that CNN based model can perform better than traditional ML models. The proposed 2D U-Net model outperformed all those models in the measurement of accuracy, precision, recall, specificity, and dice store from the table \ref{t4}, \ref{t5}. Thus, it also can be concluded that the 2D U-Net model is better for brain tumor segmentation compared to other CNN models. This experiment shows that, CNN based model are better compared to the traditional ML model and among the CNN based model, U-Net can give the precise and best result for brain tumor segmentation. The 2D U-Net model outperformed the 3d CNN based non U-Net model such as 3D FCNN. In general, 3D models are consumed more resources and time compare to 2D models. Thus, it is efficient to use 2D U-Net model instead of other 3d CNN based non U-Net model for brain tumor segmentation.

\section{Conclusion \& Future Work}
Brain tumors cause thousands of innocent victims to die due to failed brain surgery every year. However, thousands of individuals will have more hope knowing their operation will be successful if a model is developed to segment the data correctly. This study presents an extensive comparative analysis among the benchmark Brain Tumor Dataset and state-of-the-art models to predict brain tumor segmentation to achieve this goal. BraTS 2017, 2018, 2019, and 2020 datasets are used for datasets. From our analysis, we concluded that the datasets of 2017-2020 do not have many dissimilarities. So, if any models work well on one dataset, it will likely produce a good result for others. For segmentation, different CNN or modified CNN models yielded better results than traditional machine learning models. Whereas UNet produced a better score than any other CNN model. This work employed 2D UNet to predict the tumor and segment it with a 0.8409 DICE score using the 2019 BraTS dataset. 2D UNet loses information due to its inability to exploit in-depth information. In order to reduce this information loss, we aim to explore different kinds of 3D UNet models.


\bibliographystyle{IEEEtran}
\bibliography{References}

\begin{thebibliography}{10}
\providecommand{\url}[1]{#1}
\csname url@samestyle\endcsname
\providecommand{\newblock}{\relax}
\providecommand{\bibinfo}[2]{#2}
\providecommand{\BIBentrySTDinterwordspacing}{\spaceskip=0pt\relax}
\providecommand{\BIBentryALTinterwordstretchfactor}{4}
\providecommand{\BIBentryALTinterwordspacing}{\spaceskip=\fontdimen2\font plus
\BIBentryALTinterwordstretchfactor\fontdimen3\font minus
  \fontdimen4\font\relax}
\providecommand{\BIBforeignlanguage}[2]{{%
\expandafter\ifx\csname l@#1\endcsname\relax
\typeout{** WARNING: IEEEtran.bst: No hyphenation pattern has been}%
\typeout{** loaded for the language `#1'. Using the pattern for}%
\typeout{** the default language instead.}%
\else
\language=\csname l@#1\endcsname
\fi
#2}}
\providecommand{\BIBdecl}{\relax}
\BIBdecl

\bibitem{du2020medical}
G.~Du, X.~Cao, J.~Liang, X.~Chen, and Y.~Zhan, ``Medical image segmentation
  based on u-net: A review,'' \emph{Journal of Imaging Science and Technology},
  vol.~64, pp. 1--12, 2020.

\bibitem{bauer2013survey}
S.~Bauer, R.~Wiest, L.-P. Nolte, and M.~Reyes, ``A survey of mri-based medical
  image analysis for brain tumor studies,'' \emph{Physics in Medicine \&
  Biology}, vol.~58, no.~13, p. R97, 2013.

\bibitem{al2020deep}
S.~Al-Qazzaz, ``Deep learning-based brain tumour image segmentation and its
  extension to stroke lesion segmentation,'' Ph.D. dissertation, Cardiff
  University, 2020.

\bibitem{raza2022dresu}
R.~Raza, U.~I. Bajwa, Y.~Mehmood, M.~W. Anwar, and M.~H. Jamal, ``dresu-net: 3d
  deep residual u-net based brain tumor segmentation from multimodal mri,''
  \emph{Biomedical Signal Processing and Control}, p. 103861, 2022.

\bibitem{pereira2016brain}
S.~Pereira, A.~Pinto, V.~Alves, and C.~A. Silva, ``Brain tumor segmentation
  using convolutional neural networks in mri images,'' \emph{IEEE transactions
  on medical imaging}, vol.~35, no.~5, pp. 1240--1251, 2016.

\bibitem{nasim2021prominence}
M.~Nasim, A.~Dhali, F.~Afrin, N.~T. Zaman, and N.~Karim, ``The prominence of
  artificial intelligence in covid-19,'' \emph{arXiv preprint
  arXiv:2111.09537}, 2021.

\bibitem{abdullah2021prominence}
M.~Abdullah Al~Nasim, A.~Dhali, F.~Afrin, N.~T. Zaman, and N.~Karim, ``The
  prominence of artificial intelligence in covid-19,'' \emph{arXiv e-prints},
  pp. arXiv--2111, 2021.

\bibitem{palash2021fine}
M.~A.~H. Palash, M.~A. Al~Nasim, A.~Dhali, and F.~Afrin, ``Fine-grained image
  generation from bangla text description using attentional generative
  adversarial network,'' in \emph{2021 IEEE International Conference on
  Robotics, Automation, Artificial-Intelligence and Internet-of-Things
  (RAAICON)}.\hskip 1em plus 0.5em minus 0.4em\relax IEEE, 2021, pp. 79--84.

\bibitem{karim2019rl}
N.~Karim, A.~Zaeemzadeh, and N.~Rahnavard, ``Rl-ncs: Reinforcement learning
  based data-driven approach for nonuniform compressed sensing,'' in \emph{2019
  IEEE 29th International Workshop on Machine Learning for Signal Processing
  (MLSP)}.\hskip 1em plus 0.5em minus 0.4em\relax IEEE, 2019, pp. 1--6.

\bibitem{karim2021spi}
N.~Karim and N.~Rahnavard, ``Spi-gan: Towards single-pixel imaging through
  generative adversarial network,'' \emph{arXiv preprint arXiv:2107.01330},
  2021.

\bibitem{cciccek20163d}
{\"O}.~{\c{C}}i{\c{c}}ek, A.~Abdulkadir, S.~S. Lienkamp, T.~Brox, and
  O.~Ronneberger, ``3d u-net: learning dense volumetric segmentation from
  sparse annotation,'' in \emph{International conference on medical image
  computing and computer-assisted intervention}.\hskip 1em plus 0.5em minus
  0.4em\relax Springer, 2016, pp. 424--432.

\bibitem{havaei2017brain}
M.~Havaei, A.~Davy, D.~Warde-Farley, A.~Biard, A.~Courville, Y.~Bengio, C.~Pal,
  P.-M. Jodoin, and H.~Larochelle, ``Brain tumor segmentation with deep neural
  networks,'' \emph{Medical image analysis}, vol.~35, pp. 18--31, 2017.

\bibitem{menze2014multimodal}
B.~H. Menze, A.~Jakab, S.~Bauer, J.~Kalpathy-Cramer, K.~Farahani, J.~Kirby,
  Y.~Burren, N.~Porz, J.~Slotboom, R.~Wiest \emph{et~al.}, ``The multimodal
  brain tumor image segmentation benchmark (brats),'' \emph{IEEE transactions
  on medical imaging}, vol.~34, no.~10, pp. 1993--2024, 2014.

\bibitem{agravat20203d}
R.~R. Agravat and M.~S. Raval, ``3d semantic segmentation of brain tumor for
  overall survival prediction,'' in \emph{International MICCAI Brainlesion
  Workshop}.\hskip 1em plus 0.5em minus 0.4em\relax Springer, 2020, pp.
  215--227.

\bibitem{anand2020brain}
V.~K. Anand, S.~Grampurohit, P.~Aurangabadkar, A.~Kori, M.~Khened, R.~S. Bhat,
  and G.~Krishnamurthi, ``Brain tumor segmentation and survival prediction
  using automatic hard mining in 3d cnn architecture,'' in \emph{International
  MICCAI Brainlesion Workshop}.\hskip 1em plus 0.5em minus 0.4em\relax
  Springer, 2020, pp. 310--319.

\bibitem{gomathi2022glioma}
M.~Gomathi and D.~Dhanasekaran, ``Glioma detection and segmentation using deep
  learning architectures,'' \emph{Mathematical Statistician and Engineering
  Applications}, vol.~71, no.~4, pp. 452--461, 2022.

\bibitem{singh2022novel}
S.~Singh, ``A novel mask r-cnn model to segment heterogeneous brain tumors
  through image subtraction,'' \emph{arXiv preprint arXiv:2204.01201}, 2022.

\bibitem{zhuge2020automated}
Y.~Zhuge, H.~Ning, P.~Mathen, J.~Y. Cheng, A.~V. Krauze, K.~Camphausen, and
  R.~W. Miller, ``Automated glioma grading on conventional mri images using
  deep convolutional neural networks,'' \emph{Medical physics}, vol.~47, no.~7,
  pp. 3044--3053, 2020.

\bibitem{khairandish2020hybrid}
M.~O. Khairandish, R.~Gurta, and M.~Sharma, ``A hybrid model of faster r-cnn
  and svm for tumor detection and classification of mri brain images,''
  \emph{Int. J. Mech. Prod. Eng. Res. Dev}, vol.~10, no.~3, pp. 6863--6876,
  2020.

\bibitem{xu2019lstm}
F.~Xu, H.~Ma, J.~Sun, R.~Wu, X.~Liu, and Y.~Kong, ``Lstm multi-modal unet for
  brain tumor segmentation,'' in \emph{2019 IEEE 4th international conference
  on image, vision and computing (ICIVC)}.\hskip 1em plus 0.5em minus
  0.4em\relax IEEE, 2019, pp. 236--240.

\bibitem{shah2019brain}
F.~M. Shah, T.~Hossain, M.~Ashraf, F.~S. Shishir, M.~A. Al~Nasim, and M.~H.
  Kabir, ``Brain tumor segmentation techniques on medical images-a review,''
  \emph{International Journal of Scientific \& Engineering Research}, vol.~10,
  no.~2, pp. 1514--1525, 2019.

\bibitem{hossain2019brain}
T.~Hossain, F.~S. Shishir, M.~Ashraf, M.~A. Al~Nasim, and F.~M. Shah, ``Brain
  tumor detection using convolutional neural network,'' in \emph{2019 1st
  international conference on advances in science, engineering and robotics
  technology (ICASERT)}.\hskip 1em plus 0.5em minus 0.4em\relax IEEE, 2019, pp.
  1--6.

\bibitem{dong2017automatic}
H.~Dong, G.~Yang, F.~Liu, Y.~Mo, and Y.~Guo, ``Automatic brain tumor detection
  and segmentation using u-net based fully convolutional networks,'' in
  \emph{annual conference on medical image understanding and analysis}.\hskip
  1em plus 0.5em minus 0.4em\relax Springer, 2017, pp. 506--517.

\bibitem{islam2019brain}
M.~Islam, V.~Vibashan, V.~Jose, N.~Wijethilake, U.~Utkarsh, and H.~Ren, ``Brain
  tumor segmentation and survival prediction using 3d attention unet,'' in
  \emph{International MICCAI Brainlesion Workshop}.\hskip 1em plus 0.5em minus
  0.4em\relax Springer, 2019, pp. 262--272.

\bibitem{bakas2018identifying}
S.~Bakas, M.~Reyes, A.~Jakab, S.~Bauer, M.~Rempfler, A.~Crimi, R.~T. Shinohara,
  C.~Berger, S.~M. Ha, M.~Rozycki \emph{et~al.}, ``Identifying the best machine
  learning algorithms for brain tumor segmentation, progression assessment, and
  overall survival prediction in the brats challenge,'' \emph{arXiv preprint
  arXiv:1811.02629}, 2018.

\bibitem{ahmad2020context}
P.~Ahmad, S.~Qamar, L.~Shen, and A.~Saeed, ``Context aware 3d unet for brain
  tumor segmentation,'' in \emph{International MICCAI Brainlesion
  Workshop}.\hskip 1em plus 0.5em minus 0.4em\relax Springer, 2020, pp.
  207--218.

\bibitem{bauer2011fully}
S.~Bauer, L.-P. Nolte, and M.~Reyes, ``Fully automatic segmentation of brain
  tumor images using support vector machine classification in combination with
  hierarchical conditional random field regularization,'' in
  \emph{International conference on medical image computing and
  computer-assisted intervention}.\hskip 1em plus 0.5em minus 0.4em\relax
  Springer, 2011, pp. 354--361.

\bibitem{bauer2012segmentation}
S.~Bauer, T.~Fejes, J.~Slotboom, R.~Wiest, L.-P. Nolte, and M.~Reyes,
  ``Segmentation of brain tumor images based on integrated hierarchical
  classification and regularization,'' in \emph{MICCAI BraTS Workshop. Nice:
  Miccai Society}, vol.~11, 2012.

\bibitem{zikic2014segmentation}
D.~Zikic, Y.~Ioannou, M.~Brown, and A.~Criminisi, ``Segmentation of brain tumor
  tissues with convolutional neural networks,'' \emph{Proceedings
  MICCAI-BRATS}, vol.~36, no. 2014, pp. 36--39, 2014.

\bibitem{kamnitsas2016deepmedic}
K.~Kamnitsas, E.~Ferrante, S.~Parisot, C.~Ledig, A.~V. Nori, A.~Criminisi,
  D.~Rueckert, and B.~Glocker, ``Deepmedic for brain tumor segmentation,'' in
  \emph{International workshop on Brainlesion: Glioma, multiple sclerosis,
  stroke and traumatic brain injuries}.\hskip 1em plus 0.5em minus 0.4em\relax
  Springer, 2016, pp. 138--149.

\end{thebibliography}

\end{document}